\documentclass[a4paper,12pt]{article}

\pdfoutput=1

\usepackage{amsmath}
\usepackage{amssymb}
\usepackage{graphicx}
\usepackage{hyperref}
\usepackage{cite}
\usepackage{color}

\makeatletter
\@addtoreset{equation}{section}
\renewcommand{\theequation}{\thesection.\@arabic\c@equation}
\makeatother

\newcommand{\tr}{\mathrm{tr}}

\begin{document}

\begin{titlepage}

\vspace*{-15mm}   
\baselineskip 10pt   
\begin{flushright}   
\begin{tabular}{r}    
{\tt KUNS-2898}\\   
\end{tabular}   
\end{flushright}   
\baselineskip 24pt   
\vglue 10mm   

\begin{center}
{\Large\bf
 Entanglement entropy and vacuum states in Schwarzschild geometry
}

\vspace{8mm}   

\baselineskip 18pt   

\renewcommand{\thefootnote}{\fnsymbol{footnote}}

Yoshinori Matsuo\footnote[2]{ymatsuo@gauge.scphys.kyoto-u.ac.jp}, 

\renewcommand{\thefootnote}{\arabic{footnote}}
 
\vspace{5mm}   

{\it  
 Department of Physics, Kyoto University,\\Kitashirakawa, Kyoto 606-8502, Japan
}
  
\vspace{10mm}   

\end{center}

\begin{abstract}
Recently, it was proposed that there must be either large violation of the additivity conjecture 
or a set of disentangled states of the black hole in the AdS/CFT correspondence. 
In this paper, we study the additivity conjecture for 
quantum states of fields around the Schwarzschild black hole. 
In the eternal Schwarzschild spacetime, 
the entanglement entropy of the Hawking radiation is calculated assuming that 
the vacuum state is the Hartle-Hawking vacuum. 
In the additivity conjecture, we need to consider the state 
which gives minimal output entropy of a quantum channel. 
The Hartle-Hawking vacuum state does not give the minimal output entropy 
which is consistent with the additivity conjecture. 
We study the entanglement entropy in other static vacua and 
show that it is consistent with the additivity conjecture. 
\end{abstract}

\baselineskip 18pt   

\end{titlepage}

\newpage

\baselineskip 18pt

\tableofcontents


\section{Introduction}\label{sec:Intro}

Recently, it was argued that there would be either large violations of the additivity conjecture 
in quantum information theory or states without the entanglement between 
both sides of the Einstein-Rosen bridge of the black hole spacetime \cite{Hayden:2020vyo}. 
There are several statements which are equivalent to the additivity conjecture \cite{AC1,AC2,AC3,AC4,AC5,AC6,AC7}. 
A simplest statement is as follows \cite{AC3}. 
We consider two sets of density matrices $\mathcal A_\pm$ and quantum channels $\mathcal N_\pm$ 
which map density matrices in $\mathcal A_\pm$ to those in another Hilbert space $\mathcal B_\pm$. 
Then, the conjecture states that the minimum output entropy 
\begin{equation}
 S_\text{min}(\mathcal N) = \min_{\rho\in \mathcal A}S(\mathcal N(\rho)) 
\end{equation}
of two quantum channels $\mathcal N_\pm$ satisfies 
\begin{equation}
 S_\text{min}(\mathcal N_+ \otimes \mathcal N_-) = S_\text{min}(\mathcal N_+) + S_\text{min}(\mathcal N_-) \ , 
 \label{Additivity}
\end{equation}
where $S(\rho)$ is the von Neumann entropy of the state $\rho$. 
Although it is known that the conjecture is false, only small violation of the condition \eqref{Additivity} are found so far. 

In \cite{Hayden:2020vyo}, the additivity conjecture in holography was studied. 
They considered a static black hole in the asymptotic AdS spacetime. 
The geometry has two exteriors of the event horizon and two boundaries. 
The geometry corresponds to two CFTs and the states can be interpreted as the thermofield double state, 
\begin{equation}
 |\psi\rangle = \sum e^{-\beta E_n/2} |n_+\rangle |n_-\rangle \ . 
 \label{TFD}
\end{equation}
Then, we identify total systems of two CFTs to $\mathcal A_\pm$, 
subsystems of two CFTs (in subregions of the boundaries) to $\mathcal B_\pm$, 
where $\mathcal A_\pm = \mathcal B_\pm \cup \overline{\mathcal B}_\pm$, 
and the partial trace on $\overline{\mathcal B}_\pm$ to $\mathcal N_\pm$. 
Then, the entanglement entropy can be calculated by using the Ryu-Takayanagi formula 
\cite{Ryu:2006bv,Hubeny:2007xt}
and satisfies 
\begin{equation}
 S(\mathcal N_+ \otimes \mathcal N_-(\psi)) < S(\mathcal N_+(\psi)) + S(\mathcal N_-(\psi)) \ , 
\end{equation}
since the minimal surface for $\mathcal B_+$ (or $\mathcal B_-$) lies outside the horizon while 
the minimal surface for $\mathcal B_+\cup \mathcal B_-$ extends between two boundaries, and 
the area of the minimal surface for $\mathcal B_+\cup \mathcal B_-$ is smaller than twice of the area of the minimal surface for $\mathcal B_+$ \cite{Hartman:2013qma}.%
\footnote{%
The area of surface which extends between two boundaries increases with time, 
and disconnected two surfaces in each side of the horizon have smaller area at late times. 
However, it does not give the minimal output entropy as 
the entanglement entropy increases with time and becomes larger than the initial state.} 

This apparent contradiction to the additivity conjecture is obviously because 
we considered only the typical state of the black hole in the AdS spacetime. 
Assuming that the additivity conjecture is not violated in this holographic setup, 
there should be states without entanglement between 
two CFTs in the set of states after the mapping by $\mathcal N_\pm$. 
In the AdS side, disentanglement implies that 
two exteriors of the event horizon become disconnected with each other by some mechanism, 
and a few candidates were proposed in \cite{Hayden:2020vyo}. 

In this paper, we study the additivity conjecture for states in black hole spacetimes. 
The static black hole spacetime has two causally disconnected exteriors of the event horizon, 
and we identify the total systems of matter fields 
in each exterior of the event horizon to $\mathcal A_\pm$. 
We separate the total system into the Hawking radiation and the black hole, 
where the state of the Hawking radiation is identified with the state 
in the region $R$ in two exteriors of the event horizon. 
The quantum channel $\mathcal N_\pm$ projects the total system to 
the Hawking radiation by taking the partial trace in each exterior. 
Then, the set of states $\mathcal B_\pm$ is now the Hawking radiation. 
We will also consider the map to the black hole. 

In the static black hole spacetime, we usually consider only the special vacuum state 
which is known as the Hartle-Hawking vacuum \cite{Hartle:1976tp}. 
The radiation in the Hartle-Hawking vacuum is stationary --- 
incoming and outgoing radiations are balanced with each other. 
The entanglement entropy of the region $R$ increases with time 
due to the entanglement between two exteriors in the Hartle-Hawking vacuum. 
When the entanglement entropy reaches (twice of) the Bekenstein-Hawking entropy, 
the island appears and the entanglement entropy stops increasing 
\cite{Penington:2019npb,Almheiri:2019psf,Almheiri:2019hni,Almheiri:2019yqk,Penington:2019kki,Almheiri:2019qdq}.%
\footnote{
See \cite{Almheiri:2020cfm} for a review and 
\cite{Akers:2019nfi,Chen:2019uhq,Almheiri:2019psy,Chen:2019iro,Akers:2019lzs,Liu:2020gnp,Marolf:2020xie,Balasubramanian:2020hfs,
Bhattacharya:2020ymw,Verlinde:2020upt,Chen:2020wiq,Gautason:2020tmk,Anegawa:2020ezn,Hashimoto:2020cas,
Sully:2020pza,Hartman:2020swn,Hollowood:2020cou,Krishnan:2020oun,Alishahiha:2020qza,Banks:2020zrt,Geng:2020qvw,
Chen:2020uac,Chandrasekaran:2020qtn,Li:2020ceg,Bak:2020enw,Bousso:2020kmy,Anous:2020lka,Dong:2020uxp,
Krishnan:2020fer,Hollowood:2020kvk,Engelhardt:2020qpv,Karlsson:2020uga,Chen:2020jvn,Chen:2020tes,Hartman:2020khs,
Liu:2020jsv,Murdia:2020iac,Akers:2020pmf,Balasubramanian:2020xqf,Balasubramanian:2020coy,Sybesma:2020fxg,
Stanford:2020wkf,Chen:2020hmv,Ling:2020laa,Marolf:2020rpm,Harlow:2020bee,Akal:2020ujg,Hernandez:2020nem,Chen:2020ojn,Matsuo:2020ypv,Goto:2020wnk,Hsin:2020mfa,Akal:2020twv,Colin-Ellerin:2020mva,KumarBasak:2020ams,Geng:2020fxl,Karananas:2020fwx,Wang:2021woy,Marolf:2021kjc,Bousso:2021sji,Geng:2021wcq,Ghosh:2021axl,Wang:2021mqq,Kawabata:2021vyo,Akal:2021foz,Omiya:2021olc,Geng:2021hlu,Stanford:2021bhl,Balasubramanian:2021xcm,Akers:2021lms,Miyaji:2021lcq,He:2021mst,Dong:2021oad} 
for related works. 
}

This behavior of the entanglement entropy of the Hawking radiation 
is very similar to the entanglement entropy of a region in CFT in \cite{Hartman:2013qma}. 
In order to see the similarity, it is convenient to consider 
the complement of the Hawking radiation: the black hole. 
At the beginning, the region of the black hole extends between two exteriors through the horizon. 
After the Page time \cite{Page:1993wv,Page:2013dx}, 
the island appears and the region of the black hole is separated into regions in each exteriors 
\cite{Penington:2019npb,Almheiri:2019psf,Penington:2019kki,Almheiri:2019qdq}. 
This is analogous to the Ryu-Takayanagi surface in \cite{Hartman:2013qma} --- 
it initially extends between two boundaries but disconnected after some time. 
Thus, the entanglement entropy of the Hawking radiation does not give 
the minimal output entropy of the additivity conjecture in a similar fashion to \cite{Hayden:2020vyo}. 
This would be because we consider only the Hartle-Hawking state. 
Thus, we need to study the other vacuum states.

In this paper, we focus on the static vacuum states which preserve 
the static nature of the black hole geometry. 
If the amount of the stationary radiation in a static vacuum state 
is different from the Hawking radiation even slightly, 
the energy-momentum tensor diverges at the event horizon in the classical black hole geometry. 
This divergence implies that quantum corrections 
to the energy-momentum tensor in the Einstein equation are no longer negligible. 
The black hole geometry is modified from the classical solution by quantum effects of the vacuum states. 
By taking this quantum effect into account, 
solutions of the semi-classical Einstein equation in the other vacuum than the Hartle-Hawking vacuum 
have no event horizon and the two exteriors of the event horizon are disconnected. 
Then, the entanglement entropy is given by sum of two exteriors, 
and the additivity conjecture is satisfied. 

This paper is organized as follows. 
In Sec.~\ref{sec:Green}, we review on the vacuum state, 
the entanglement entropy and the energy-momentum tensor in the curved spacetime. 
We discuss how the entanglement entropy and the energy-momentum tensor depend on the vacuum states. 
In Sec.~\ref{sec:Schwarz}, we study the entanglement entropy 
in static vacua around the Schwarzschild black hole. 
Sec.~\ref{sec:Conclusion} is devoted for the conclusion and discussions.


\section{Entanglement entropy and vacuum state}\label{sec:Green}

In this section, we briefly review basic facts on 
the entanglement entropy and vacuum states, 
and then, see how the entanglement entropy depends on the vacuum states. 
Here, we focus on two-dimensional massless fields for simplicity. 
The vacuum state has the lowest energy and 
depends on the choice of the associated time coordinate. 
Although the entanglement entropy is invariant under the coordinate transformation, 
it depends 
on the coordinates which is used to define the vacuum state. 
The entanglement entropy is sometimes 
calculated without specifying the vacuum state explicitly. 
However, the state is determined implicitly when we choose coordinates in the concrete calculation. 
Here, we will see the relation explicitly. 
We will also study the relation to the energy-momentum tensor.


\subsection{Vacuum state}\label{sec:Vacuum}

We consider a two-dimensional massless free scalar field $\phi$. 
The action is given by  
\begin{equation}
 S = - \frac{1}{2}\int d^2x \sqrt{-g} g^{\mu\nu} \partial_\mu \phi \partial_\nu \phi 
 = - \frac{1}{2}\int du\,dv\,\partial_u \phi \partial_v \phi \ , 
\end{equation}
where $u$ and $v$ are the null coordinates. 
Since the action does not depend on the metric, 
the scalar field can always be expanded in the Fourier modes. 
In the coordinates $(u,v)$, the Fourier expansion is given as 
\begin{equation}
 \phi(x) 
 = 
 \int_0^\infty \frac{ d \omega}{2\pi} \frac{1}{\sqrt{2\omega}} 
 \left[a_\omega e^{-i\omega v} + a_\omega^\dag e^{i\omega v}
 + b_\omega e^{-i\omega u} + b_\omega^\dag e^{i\omega u}\right] \ , 
\end{equation}
where $a_\omega$ and $a_\omega^\dag$ are annihilation and creation operators 
of the incoming modes and $b_\omega$ and $b_\omega^\dag$ are those of outgoing modes, respectively. 
The vacuum state $|0\rangle$ is defined as the state which is annihilated by the annihilation operators, 
\begin{align}
 a_\omega |0\rangle &= 0 \ , 
 &
 b_\omega |0\rangle &= 0 \ . 
\end{align}
The vacuum state $|0\rangle$ is associated to the coordinates $(u,v)$, 
since the annihilation and creation operators are defined by using the coordinates $(u,v)$. 
Another vacuum state $|\Omega\rangle$ can be also defined by using another pair of coordinates $(U,V)$. 
The creation and annihilation operators are defined in a similar fashion as 
\begin{equation}
 \phi(x) 
 = 
 \int_0^\infty \frac{ d \omega}{2\pi} \frac{1}{\sqrt{2\omega}} 
 \left[\tilde a_\omega e^{-i\omega V} + \tilde a_\omega^\dag e^{i\omega V}
 + \tilde b_\omega e^{-i\omega U} + \tilde b_\omega^\dag e^{i\omega U}\right] \ , 
\end{equation}
and then, the vacuum state $|\Omega\rangle$ is defined by 
\begin{align}
 \tilde a_\omega |\Omega\rangle &= 0 \ , 
 &
 \tilde b_\omega |\Omega\rangle &= 0 \ , 
\end{align}
but, in general, is not annihilated by $a_\omega$ or $b_\omega$; 
\begin{align}
 a_\omega |\Omega\rangle &\neq 0 \ , 
 &
 b_\omega |\Omega\rangle &\neq 0 \ , 
\end{align}
since two different annihilation operators are related to each other as 
\begin{equation}
 a_\omega 
 = 
 \int d \omega' 
 \left(A_{\omega\omega'} \tilde a_{\omega'} + A_{\omega,-\omega'} \tilde a_{\omega'}^\dag \right) \ , 
\end{equation}
where 
\begin{equation}
 A_{\omega\omega'} = \sqrt{ \frac{\omega}{\omega'}} \int dv\, e^{i \omega v}e^{-i \omega' V(v)} \ . 
\end{equation}

The correlation functions in the vacuum state $|\Omega\rangle$ is different from those in $|0\rangle$. 
It is straightforward to calculate the two-point correlation function in the vacuum state $|0\rangle$ as  
\begin{align}
 \langle 0 | \phi(x) \phi(x') |0\rangle 
 &= 
 - \int \frac{ d \omega}{4\pi \omega} \left[e^{-i\omega(v-v')} + e^{-i\omega(u-u')}\right]
 \notag\\
 &= - \frac{1}{4\pi}\log \left|(u-u')(v-v')\right| \ . 
 \label{Green-0}
\end{align}
The two-point function in $|\Omega\rangle$ is calculated in a similar fashion, 
but has a different form from the one in $|0\rangle$; 
\begin{align}
 \langle \Omega | \phi(x) \phi(x') |\Omega \rangle 
 &= - \frac{1}{4\pi}\log \left|(U-U')(V-V')\right| 
 \neq
 \langle 0 | \phi(x) \phi(x') |0\rangle \ . 
 \label{Green-O}
\end{align}
In general, the two-point function satisfies 
\begin{equation}
 \partial_u \partial_v \left\langle \phi(x) \phi(x') \right\rangle 
 = \delta^{(2)}(x-x') \ , 
\end{equation}
where $u$ and $v$ are the null coordinates at $x$. 
The correlation function is universal up to regular terms, 
\begin{equation}
 \left\langle \phi(x) \phi(x') \right\rangle 
 = - \frac{1}{4\pi}\log \left|(u-u')(v-v')\right| + \text{(regular terms)} \ . 
 \label{Green}
\end{equation}
Thus, the difference between the correlation functions 
in different vacua, \eqref{Green-0} and \eqref{Green-O}, 
is nothing but the non-universal regular terms. 
The correlation function in general has 
the additional regular terms which depend on the state, 
but has no additional terms 
if it is written in terms of the coordinates 
associated to the vacuum state.


\subsection{Entanglement entropy}\label{sec:Entropy}

The entanglement entropy can be calculated by using the replica method 
\cite{Callan:1994py,Holzhey:1994we,Calabrese:2004eu,Casini:2005rm,Calabrese:2009qy}. 
The entanglement entropy $S$ is expressed in terms of the density matrix $\rho$ as 
\begin{equation}
 S = -\tr\rho \log \rho = \lim_{n\to 1} \frac{1}{1-n} \log\tr\rho^n \ .
\end{equation}
For the density matrix of a region $A$ in a two-dimensional field theory, $\tr\rho^n$ is given in terms of 
the partition function on $n$-sheeted Riemann surface $Z_n$ as, 
\begin{equation}
 \tr\rho^n = \frac{Z_n}{Z_1} \ , 
\end{equation}
where the Riemann surface has the branch cut(s) on the region $A$, 
and each sheet of the Riemann surface is a copy of the original spacetime. 
The partition function $Z_n$ can be obtained by introducing the twist operators 
\begin{equation}
 Z_n = \prod_k \left\langle \prod_i e^{\frac{k}{n}\varphi(x_i)} e^{-\frac{k}{n}\varphi(x'_i)} \right\rangle \ , 
\end{equation}
where $e^{\pm\frac{k}{n}\varphi}$ are the twist operators on the $k$-th sheet 
and the sign comes from the direction of the twist. 
The endpoints of the branch cut(s) are located at $x_i$ and $x'_i$. 

If the vacuum state is associated to the coordinates $(u,v)$, 
the correlation function of $\varphi$ is given by%
\footnote{%
It should be noted that the field $\varphi$ is not identical to matter fields $\phi$
but the correlation function is related to that of $\phi$. 
The twist operators can be expressed in terms of some current operators associated to each matter field \cite{Casini:2005rm}, 
and hence, the correlation function of $\varphi$ has the additional regular terms 
if the correlation function of $\phi$ has the additional regular terms. 
} 
\begin{equation}
 \langle\varphi(x)\varphi(x')\rangle = \log\left|(u-u')(v-v')\right| \ , 
\end{equation}
and then, the entanglement entropy is calculated as 
\begin{equation}
 S = \frac{c}{12}\sum_{i,j} \log\left|\frac{(u_i-u'_j)^2(v_i-v'_j)^2}{(u_i-u_j)(v_i-v_j)(u'_i-u'_j)(v'_i-v'_j)}\right| \ , 
\end{equation}
where $c$ is the central charge, or equivalently, the number of the massless fields. 
The summation in the expression above includes $i=j$, 
which come from the self-interaction of the twist operators and give the UV divergence. 
The divergence should be regularized by introducing a cut-off, 
\begin{align}
 u_i-u_i  &\to \epsilon^u \ , 
 &
 v_i - v_i & \to \epsilon^v \ .  
\end{align}
In the case of the curved spacetime, the cut-off should be given in the proper distance
\begin{equation}
 \epsilon^2 = 2 g_{uv} \epsilon^u \epsilon^v \ , 
\end{equation}
and then, the entanglement entropy is expressed as 
\begin{align}
 S 
 &= 
 \frac{c}{6}\sum_{i,j}\log\left|(u_i-u'_j)(v_i-v'_j)\right|
 + \frac{c}{12} \sum_i \log\left|\frac{4g_{uv}(x_i)g_{uv}(x'_i)}{\epsilon^4}\right|
 \notag\\
 &\quad
 - \frac{c}{6}\sum_{i<j}\log\left|(u_i-u_j)(v_i-v_j)\right|
 - \frac{c}{6}\sum_{i<j}\log\left|(u'_i-u'_j)(v'_i-v'_j)\right|
 \ . 
 \label{S2D}
\end{align}
As we discussed, the coordinates in the expression above should be chosen so that 
it agrees with the coordinate which is used to define the vacuum state. 
The expression above also contains the metric component, 
which is also in the same coordinates. 
Note that the expression above is not given in terms of the proper distance. 
The same expression with different coordinate gives a different value of the entanglement entropy, 
which is nothing but the vacuum dependence of the entanglement entropy. 

In this section, we assumed that there is no boundary within a finite distance 
from the twist operators, and the number of the twist operators are even. 
In the presence of the boundary, the entanglement entropy can be calculated 
by introducing the mirror image on the other side of the boundary, 
and then, the number of the twist operators also becomes even, effectively. 
If the number of the twist operators is odd, one of the branch cut extends to the boundary, 
and hence, we need to take the effect of the boundary 
into the calculation even if the spacetime is non-compact. 
In this case, we need to introduce a regularization 
by putting the boundary in a finite distance $\Lambda$ 
and take the limit of $\Lambda\to\infty$. 
Then, we can still use the formula \eqref{S2D} in these cases. 
If there is degrees of freedom which are localized on the boundary, 
we need to add the contribution from the boundary entropy.


\subsection{Energy-momentum tensor}\label{sec:EM tensor}


In this section, we study the energy-momentum tensor in curved spacetimes 
and see how it depends on the vacuum states. 
Before calculating the energy-momentum tensor in curved spacetimes, 
we first consider the energy-momentum tensor in the flat spacetime 
and see that the energy-momentum tensor has different expectation values 
in the different vacua $|0\rangle$ and $|\Omega\rangle$. 
The energy-momentum tensor of two-dimensional free scalar fields in the flat spacetime is given by 
\begin{equation}
 \langle T_{uu}^{(2D)}\rangle 
 = 
 \lim_{x'\to x} \langle \psi | : \partial_u\phi(x) \, \partial_{u'}\phi(x') : |\psi \rangle  \ , 
\end{equation}
and similarly for $T_{vv}$, where $|\psi\rangle$ is a quantum state. 
Here, we take $(u,v)$ to be the standard null coordinates of the flat spacetime; namely, 
the two-dimensional metric is expressed as 
\begin{equation}
 ds^2 = -du\,dv \ . 
\end{equation}
We consider the gravitational theory which has the flat spacetime 
as a solution of the semi-classical Einstein equation in the vacuum state 
which is defined by the coordinates $(u,v)$. 
The normal ordering is defined in terms of 
the creation and annihilation operators associated to the coordinates $(u,v)$, 
and then, the expectation value of the energy-momentum tensor is zero in the vacuum state $|0\rangle$. 
In the vacuum state $|\psi\rangle = |\Omega\rangle$, 
which is associated to the coordinates $(U,V)$, 
the energy-momentum tensor is calculated as 
\begin{align}
 \langle T_{uu}^{(2D)} \rangle
 &= - \frac{1}{4\pi} \lim_{u'\to u} 
 \partial_u \partial_{u'} \log\left|\frac{U(u)-U(u')}{u-u'}\right|  
 \notag\\
 &= - \frac{1}{24\pi} \left\{U,u\right\} \ , 
\end{align}
where $\{f,x\}$ is the Schwarzian derivative. 

Now, we consider the energy-momentum tensor of the s-waves in the four-dimensional spacetime. 
The metric of the four-dimensional spherically symmetric spacetime can always be written as
\begin{equation}
 ds^2 = - C(u,v) du\,dv + r^2(u,v) d \Omega^2 \ ,
 \label{metric0}
\end{equation}
where $d \Omega^2$ is the metric on a unit 2-sphere. 
In the s-wave approximation, the energy-momentum tensor is given 
in terms of the energy-momentum tensor of two-dimensional fields as 
\begin{equation}
 T_{\mu\nu} = \frac{1}{4\pi r^2} T_{\mu\nu}^{(2D)} \ , 
\end{equation}
if both $\mu$ and $\nu$ are in $(u,v)$-directions, and the other angular components vanish. 
The energy-momentum tensor of two-dimensional massless fields 
has the Weyl anomaly, 
\begin{equation}
 \langle T^\mu{}_{\mu}\rangle = \frac{c}{24\pi} R^{(2D)} \ , 
\end{equation}
where $R^{(2D)}$ is the scalar curvature of two-dimensional spacetime 
and $c$ is the number of fields. 
Together with the conservation law, the energy-momentum tensor is completely fixed 
up to integration constants. Thus, the energy-momentum tensor is given by \cite{Davies:1976ei,Davies:1976hi}
\begin{align}
\langle T_{uu} \rangle &=
- \frac{c}{48\pi^2 r^2} C^{1/2} \partial_u^2 C^{-1/2} + \frac{c\,F(u)}{96\pi^2 r^2} \ , 
\label{Tuu-vac}
\\
\langle T_{vv} \rangle &=
- \frac{c}{48\pi^2 r^2} C^{1/2} \partial_v^2 C^{-1/2} + \frac{c\,\bar F(v)}{96\pi^2 r^2} \ , 
\label{Tvv-vac}
\\
\langle T_{uv} \rangle &=
- \frac{c}{96\pi^2 r^2 C^2}
\left[ C \partial_u\partial_v C - \partial_u C \partial_v C \right],
\label{Tuv-vac}
\\
\langle T_{\theta\theta} \rangle &= 0 \ , 
\label{Tthth-vac}
\end{align}
where $F(u)$ and $\bar F(v)$ are the integration constants, 
which represent the outgoing and incoming energy. 

The energy-momentum tensor transforms covariantly under the coordinate transformation, 
but the expression above is not written in a covariant form. 
The first terms in \eqref{Tuu-vac} and \eqref{Tvv-vac} do not transform covariantly, 
but the non-covariant part is absorbed by 
the non-trivial transformation of the integration constants $F(u)$ and $\bar F(v)$. 
The above expression is valid for any physical state $|\psi\rangle$, 
and the energy-momentum tensor depends on the state $|\psi\rangle$ 
through the integration constants $F(u)$ and $\bar F(v)$. 

Suppose $|0\rangle$ and $|\Omega\rangle$ are the vacuum state associated to 
the coordinates $(u,v)$ and $(U,V)$, respectively, 
and the metric is given by 
\begin{equation}
 ds^2 = - C du\,dv + r^2 d \Omega = - \widetilde C dU dV + r^2 d \Omega \ , 
\end{equation}
where 
\begin{equation}
 \widetilde C = C 
 \left(\frac{\partial U(u)}{\partial u}\right)^{-1} \left(\frac{\partial V(v)}{\partial v}\right)^{-1} \ . 
\end{equation}
Once we fix the integration constants for a vacuum state $|0\rangle$ 
by using \eqref{Tuu-vac}--\eqref{Tthth-vac} in the $(u,v)$-coordinates, 
\eqref{Tuu-vac}--\eqref{Tthth-vac} in the $(U,V)$-coordinates 
with the same integration constants gives the expectation value in $|\Omega\rangle$. 
\begin{align}
 - & \frac{c}{48\pi^2 r^2} \left(\frac{\partial U(u)}{\partial u}\right)^2 
 \widetilde C^{1/2} \partial_U^2 \widetilde C^{-1/2} + \frac{F(u)}{96\pi^2 r^2} 
 \notag\\
 &= 
 - \frac{c}{48\pi^2 r^2} C^{1/2} \partial_u^2 C^{-1/2} + \frac{F(u)}{96\pi^2 r^2} 
 - \frac{1}{96\pi^2 r^2} \left\{U,u\right\} 
 \notag\\
 &= \langle \Omega | T_{uu} |\Omega\rangle \ , 
\end{align}
where, in the last equality, we used 
\begin{align}
 \langle \Omega | T_{uu} |\Omega\rangle - \langle 0 | T_{uu} |0\rangle 
 &= 
 - \frac{1}{16\pi^2r^2} \lim_{u'\to u} 
 \partial_u \partial_{u'} \log\left|\frac{U(u)-U(u')}{u-u'}\right|  
 \notag\\
 &= 
 - \frac{1}{96\pi^2 r^2} \left\{U,u\right\} \ ,  
 \label{EMtensorDiff}
\end{align}
since the Green function of the two-dimensional massless field does not depend on the metric, 
and the metric dependence of the energy-momentum tensor comes from 
the universal UV regularization which is independent of the physical state $|\psi\rangle$.


\section{Entanglement entropy in Schwarzschild spacetime}\label{sec:Schwarz}

In this section, we study the entanglement entropy of matter fields 
in the Schwarzschild spacetime. 
We focus on static vacuum states. 
In the Hartle-Hawking vacuum \cite{Hartle:1976tp}, 
the energy-momentum tensor is sufficiently small 
and the geometry is approximately given by the classical Schwarzschild solution. 
In this case, quantum states in the left and right exteriors of the event horizon 
are connected by the Einstein-Rosen bridge and hence are entangled with each other. 

In the other static vacua, 
the expectation value of the energy-momentum tensor diverges 
at the event horizon in the classical Schwarzschild solution. 
This implies that the effect of the energy-momentum tensor of the quantum vacuum state 
in the semi-classical Einstein equation is not negligible near the horizon. 
By taking this effect into account, the geometry is modified from 
the classical Schwarzschild solution. 
The geometry has no event horizon at finite radius 
and two exteriors of the horizon in the classical solution 
are now disconnected from each other. 
Thus, quantum states in the left and right exteriors are disentangled with each other. 

Here, we consider the Schwarzschild black hole 
in the four-dimensional asymptotically flat spacetime. 
The gravitational part of the action is given by 
the Einstein-Hilbert action with the Gibbons-Hawking term, 
\begin{align}
 I &= I_\text{gravity} + I_\text{matter} \, ,
 \\
 &I_\text{gravity} 
 = 
 \frac{1}{16\pi G_N} \int_{\mathcal M} d^4 x \sqrt{-g} \, R 
 + \frac{1}{8\pi G_N} \int_{\partial \mathcal M} d^3 x \sqrt{-h} \, K \ , 
\end{align}
where $G_N$ is the Newton constant. 

The Schwarzschild metric is given by 
\begin{equation}
 ds^2 = - \left(1 - \frac{r_h}{r}\right) dt^2 + \left(1 - \frac{r_h}{r}\right)^{-1} dr^2 + r^2 d \Omega^2 \ , 
 \label{MetricS}
\end{equation}
where $r_h$ is the Schwarzschild radius. 
The metric is rewritten in the form of \eqref{metric0} with $C(u,v)$ and $r(u,v)$ given by 
\begin{align}
 C(r_*) &= 1 - \frac{r_h}{r(r_*)} \ , 
 \label{cls-c}
 \\
 r_* &= r-r_h + r_h \log\left(\frac{r-r_h}{r_h}\right) \ , 
 \label{cls-r}
\end{align}
and the tortoise coordinates are related to the coordinates $(u,v)$ as 
\begin{align}
 u &= t - r_* \ , 
 &
 v = t + r_* \ . 
 \label{uv}
\end{align}

Now, we study the vacuum state on this metric. 
The energy-momentum tensor is given by \eqref{Tuu-vac}--\eqref{Tthth-vac} and 
the integration constants in \eqref{Tuu-vac} and \eqref{Tvv-vac} are related to the vacuum state. 
First, we fix the integration constants in so that the energy-momentum tensor vanishes in the flat spacetime. 
This vacuum state is nothing but the Boulware vacuum \cite{Boulware:1974dm} since the condition implies 
there is neither incoming nor outgoing energy in the asymptotically flat region. 
In $(u,v)$ coordinates \eqref{uv}, \eqref{cls-c} gives $C\to 1$ in the asymptotic region $r\to \infty$, 
and then, the energy-momentum tensor \eqref{Tuu-vac}--\eqref{Tthth-vac} 
vanishes except for the integration constants $F(u)$ and $\bar F(v)$. 
Thus, the energy-momentum tensor in the Boulware vacuum is given by \eqref{Tuu-vac}--\eqref{Tthth-vac}  
with $F(u) = \bar F(v) = 0$ by using the coordinates \eqref{uv}. 

The energy-momentum tensor in the other static vacuum states 
can be obtained by taking $F(u) = \bar F(v) = \text{const.}$, or equivalently, 
$F(u) = \bar F(v) = 0$ by using other coordinates, 
as we have discussed in Sec.~\ref{sec:EM tensor}. 
Static vacuum states with arbitrary amount of radiation 
can be obtained by using the coordinates 
\begin{align}
 U &= - \kappa^{-1} e^{-\kappa u} \ , 
 &
 V &= \kappa^{-1} e^{\kappa v} \ , 
 \label{UV}
\end{align}
and then, by using the expressions \eqref{Tuu-vac}--\eqref{Tthth-vac} with the coordinates \eqref{UV}, 
the energy-momentum tensor in the asymptotically flat region is given by 
\begin{align}
 T_{uu} = T_{vv} = \frac{c\,\kappa^2}{196\pi^2 r^2} + \mathcal O(r^{-3}) \ . 
\end{align}
When we use the expressions \eqref{Tuu-vac}--\eqref{Tthth-vac} in coordinates \eqref{uv}, 
the integration constants becomes 
\begin{align}
 F(u) = \bar F(v) = \frac{\kappa^2}{2} \ . 
 \label{F-vac}
\end{align}
The vacuum state is nothing but the Hartle-Hawking vacuum \cite{Hartle:1976tp} when $\kappa$ agrees with 
the surface gravity $\kappa = \frac{1}{2 r_h}$. 
The other static vacuum states correspond to other values of $\kappa$. 
The coordinates $(U,V)$ reduce to the Schwarzschild coordinates $(u,v)$ in the $\kappa\to 0$ limit. 

By using the coordinates \eqref{UV}, the semi-classical Schwarzschild metric is expressed as 
\begin{equation}
 ds^2 = - \frac{dU dV}{W^2} + r^2 d \Omega^2 \ , 
 \label{metric-UV}
\end{equation}
where the coordinates $(U,V)$ are defined by \eqref{UV}, and $W$ is defined as 
\begin{equation}
 W = \sqrt{\frac{r \kappa^2 \left|U V \right|}{r-r_h}} 
 = \kappa\sqrt{\frac{r}{r_h}}\left(\frac{r-r_h}{r_h}\right)^{\kappa r_h - 1/2} e^{\kappa(r-r_h)} \ . 
 \label{W-away}
\end{equation}

Now, we consider effects of 
the quantum energy-momentum tensor \eqref{Tuu-vac}--\eqref{Tthth-vac} to the geometry. 
By taking the quantum effects into account, 
the geometry is given by a solution of the semi-classical Einstein equation, 
\begin{equation}
 R_{\mu\nu} - \frac{1}{2} g_{\mu\nu} R = 8 \pi G_N \langle T_{\mu\nu} \rangle \ . 
 \label{SemiEin}
\end{equation}
In most cases, quantum effects are sufficiently smaller than the classical energy-momentum tensor
and the geometry can be determined by the classical Einstein equation. 
However, the energy-momentum tensor in the Schwarzschild spacetime \eqref{Tuu-vac}--\eqref{Tthth-vac} 
diverges at the event horizon in the other static vacua 
than the Hartle-Hawking vacuum, $\kappa\neq \frac{1}{2 r_h}$. 
Thus, around the Schwarzschild radius of the Schwarzschild black hole, 
quantum effects in the energy-momentum tensor is very large and 
contribute to the leading order of the semi-classical Einstein equation \eqref{SemiEin}

Near the Schwarzschild radius, 
the classical solution, \eqref{cls-c} and \eqref{cls-r}, is approximated as 
\begin{align}
 C(x) &\simeq \frac{\alpha}{4r_h^2} e^{x/r_h} \ , 
 \label{cls-c-nh}
 \\
 r(x) &\simeq r_h + \frac{\alpha}{4r_h} e^{x/r_h} \ . 
 \label{cls-r-nh}
\end{align}
where 
\begin{align}
 \alpha &= \frac{c G_N}{12\pi} \ , 
 \label{alpha}
 \\
 x &= r_* - r_h \log\left(\frac{\alpha}{4r_h^2}\right) \ . 
 \label{x}
\end{align}
By solving the semi-classical Einstein equation \eqref{SemiEin} with 
\eqref{Tuu-vac}--\eqref{Tthth-vac} near the Schwarzschild radius, we obtain 
\cite{Fabbri:2005zn,Fabbri:2005nt,Ho:2017joh,Ho:2017vgi}
\begin{align}
 C(x) &= \frac{\alpha}{4r_h^2} e^{x/r_h} + \mathcal O(\alpha^2) \ , 
 \label{qtm-c}
 \\
 r(x) &= r_h + \frac{\alpha}{4r_h} e^{x/r_h} 
 - \alpha\left(\frac{1}{4r_h^2} - \kappa^2\right) 
 \left(x + r_h \log\frac{\alpha}{4r_h^2}\right) + \mathcal O(\alpha^2) \ . 
 \label{qtm-r}
\end{align}
(See Appendix~\ref{sec:nh} for details of the derivation.)
When the metric is written in the coordinate \eqref{UV}, 
$W$ in \eqref{metric-UV} is approximated near the horizon as%
\footnote{%
Since $C$ has no significant correction, $W$ also has no important correction 
when it is written in terms of the tortoise coordinate $r_*$. 
However, the expression \eqref{W-away} is no longer valid, 
as the relation between $r$ and $r_*$ is modified 
from \eqref{cls-r} to \eqref{qtm-r} by the quantum correction. 
} 
\begin{equation}
 W = \left(\frac{\alpha}{4 r_h^2}\right)^{\kappa r_h -1/2} e^{\left(\kappa - \frac{1}{2r_h}\right)x } \ . 
 \label{W-near}
\end{equation}
 
The quantum correction near the Schwarzschild radius vanishes in 
the Hartle-Hawking vacuum, $\kappa = \frac{1}{2r_h}$. 
In other cases, the geometry has no horizon at finite radius. 
For $\kappa < \frac{1}{2r_h}$, the radius of 2-sphere $r$ 
has local minimum around the Schwarzschild radius. 
The radius becomes larger as approaches to the black hole 
beyond the local minimum and eventually goes to infinity. 
The geometry has no event horizon but reaches $r\to\infty$ within a finite proper distance. 
For $\kappa > \frac{1}{2 r_h}$, the radius decreases much faster than 
the case of the Hartle-Hawking vacuum. 
The event horizon is located in $x\to -\infty$ in the classical Schwarzschild geometry, 
but the quantum correction implies that the radius shrinks to zero 
before crossing the event horizon. 
The geometry has no event horizon but has the conical singularity at $r=0$. 
Thus, two exteriors are connected by the Einstein-Rosen bridge only in the Hartle-Hawking vacuum. 
They are disconnected in the other static vacua due to the quantum effect of the vacua. 


\subsection{Entanglement entropy}\label{sec:Island}

\begin{figure}
\begin{center}
\includegraphics[scale=0.3]{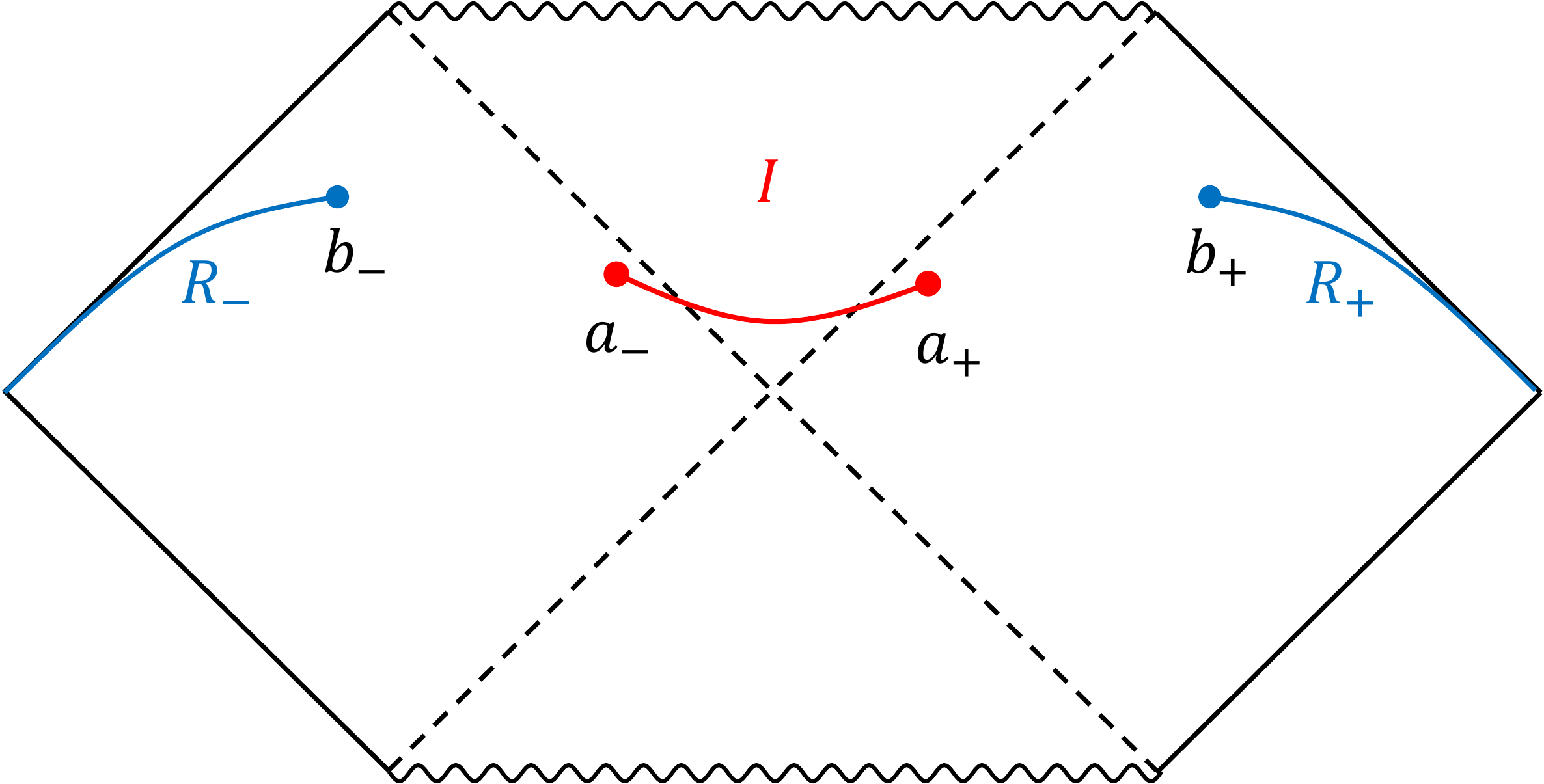}
\hspace{24pt}
\includegraphics[scale=0.3]{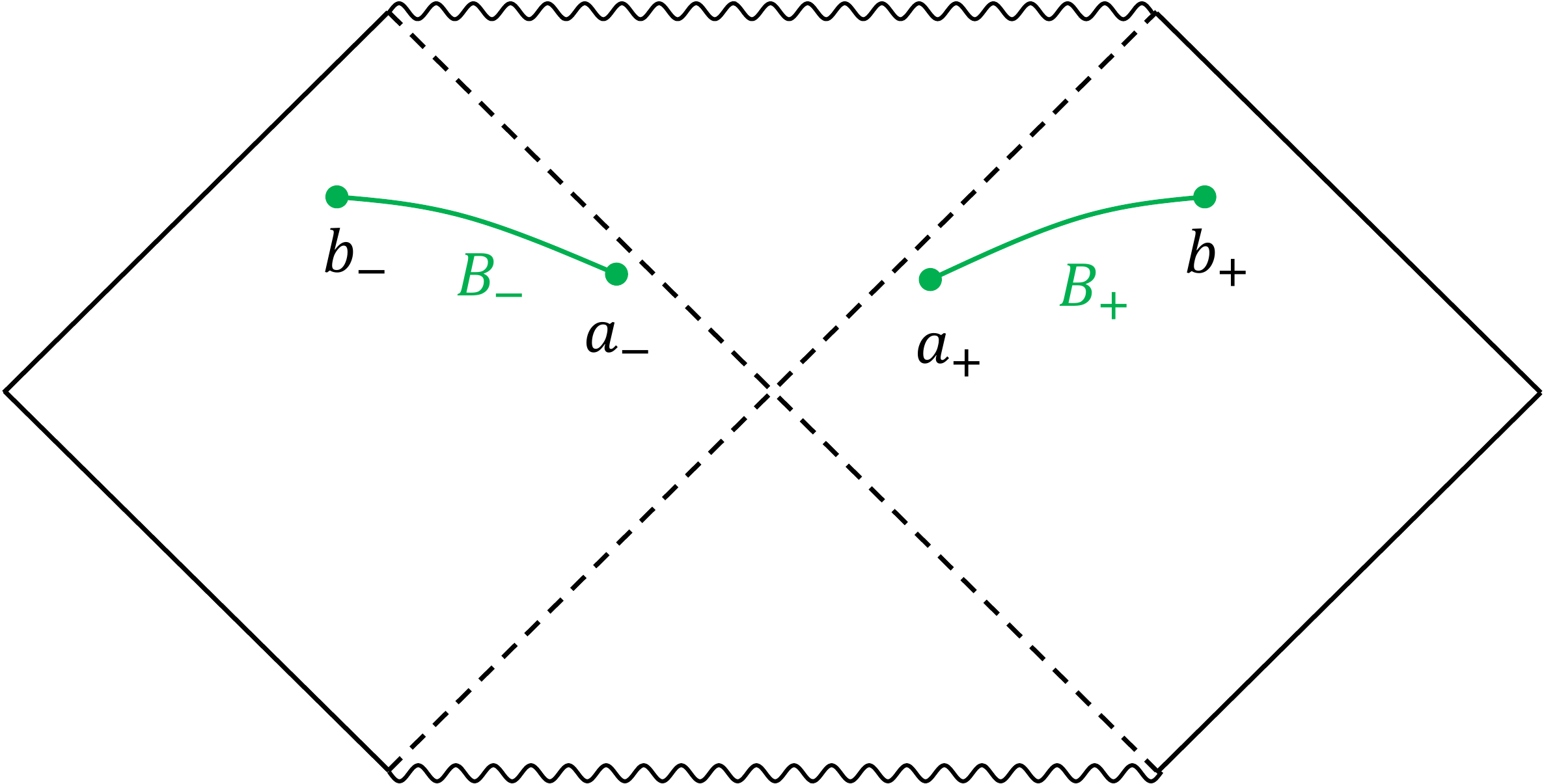}
\caption{%
We separate the total system into the Hawking radiation and black hole. 
The entanglement entropy of the Hawking radiation is identified with 
the entanglement entropy of the region $R = R_+ \cup R_-$, which is defined as the region outside $b_\pm$.  
The entanglement entropy of the black hole is given by 
the entanglement entropy of the region $B = B_+ \cup B_-$, which is defined as 
the region between $a_\pm$ and $b_\pm$. 
The position of the quantum extremal surface $a_\pm$ is determined so that 
the entanglement entropy is extremized. 
The island is effectively included in either the Hawking radiation or black hole 
due to gravitational effects in the replica geometries. 
Before the Page time, the entanglement entropy of the black hole is 
effectively given by the entanglement entropy of the reion $B\cup I$. 
After the Page time, the entanglement entropy of the Hawking radiation 
is effectively given by the entanglement entropy of the region $R\cup I$. 
}\label{fig:BH}
\end{center}
\end{figure}

In this paper, we consider the output entropy of quantum channels $\mathcal N_\pm$. 
We separate the total system into the Hawking radiation and the black hole.%
\footnote{%
Here, we separate the spacetime by the surface at a radius $b$. 
Here, the Hawking radiation stands for quantum fields in $r>b$, 
or equivalently, in $R_\pm$ (See Fig.~\ref{fig:BH}), 
though quantum fields in $R_\pm$ is not strictly only the Hawking radiation. 
The region of the black hole is $r<b$ but outside the island $I$. 
} 
The quantum channel $\mathcal N_+$ ($\mathcal N_-$) maps 
states in the total system in the right (left) exterior of the horizon to 
states of the Hawking radiation in the right (left) exterior by taking the partial trace. 
The states and the entanglement entropy of the Hawking radiation 
are identified with those in the region $R_\pm$. 
Then, the entanglement entropy is calculated by using the replica trick. 

The entanglement entropy is given by 
\cite{Faulkner:2013ana,Lewkowycz:2013nqa,Engelhardt:2014gca,Dong:2016hjy,Dong:2017xht} 
\begin{equation}
 S
 = 
 \sum_{\partial R,\ \partial I} \frac{\text{Area}}{4 G_N} 
 + \sum_{\partial R,\ \partial I} S_\text{matter}^\text{(non-local)} \ , 
 \label{IslandRule}
\end{equation}
where the area term comes from the gravitational part. 
The leading contribution from the matter part gives a similar area term 
\cite{Bombelli:1986rw,Srednicki:1993im}, 
which can be absorbed by the gravitational part as the renormalization of 
the gravitational coupling constants \cite{Susskind:1994sm}. 
In the s-wave approximation, contributions from the matter part 
is given by the formula of the entanglement entropy of two-dimensional massless fields, \eqref{S2D}. 

In the case of the gravitational theory, the dominant configuration possibly 
has replica wormholes, which connect different sheets of replicas. 
The replica wormhole gives the same contributions to branch cuts. 
Thus, the entanglement entropy has contribution from the islands, 
where the replica wormhole is located. 
Thus, the endpoints of branch cuts in the formula \eqref{S2D} 
includes those of the region $R$ and the island $I$. 

In this section, we calculate the entanglement entropy in static vacua of matter fields. 
The entanglement entropy depends on the vacuum state through the matter part. 
By taking the effect of the vacuum energy in the semi-classical Einstein equation, 
the semi-classical geometry also depends on the vacuum states. 
In the following, we will calculate the entanglement entropy 
in the Hartle-Hawking vacuum, the Boulware vacuum and the other static vacua, 
which are defined by the coordinates \eqref{UV}.


\subsubsection{Entanglement entropy in Hartle-Hawking vacuum}\label{sec:HH}

Here, we consider the entanglement entropy of the Hawking radiation in the Hartle-Hawking vacuum. 
The energy-momentum tensor in the Hartle-Hawking vacuum has no divergence, and 
the back-reaction to the geometry from the quantum vacuum energy is sufficiently small. 
Thus, the entanglement entropy is approximately given by 
the formula \eqref{IslandRule} on the classical Schwarzschild geometry. 
The entanglement entropy of $R = R_+\cup R_-$ and $R_+$ are calculated 
in \cite{Hashimoto:2020cas} and \cite{Matsuo:2020ypv}, respectively. 
In these papers, the matter part of the entanglement entropy \eqref{S2D} 
in the s-wave approximation is calculated by using the Kruskal coordinates 
and hence, the vacuum states are chosen to be the Hartle-Hawking vacuum, implicitly. 
Here, we review the calculations in these papers and see 
the entanglement entropy in the Hartle-Hawking vacuum 
does not give the minimal output entropy which satisfies the additivity conjecture, 
and has a similar behavior to the entanglement entropy in the holographic setup 
\cite{Hartman:2013qma,Hayden:2020vyo}. 

First, we review the entanglement entropy of the Hawking radiation in the region $R$, 
which is the output entropy of the quantum channel $\mathcal N_+\otimes\mathcal N_-$ 
in the additivity conjecture \eqref{Additivity}. 
The region $R$ consists of two connected regions $R_\pm$ in both two exteriors, 
and $R_\pm$ extend from the inner boundaries $b_\pm$ to the spatial infinity. 
The positions of $b_+$ and $b_-$ are given by $(t,r) = (t_b,b)$ and $(t,r) = (- t_b + i \beta/2,b)$, 
respectively, where the imaginary part of the time indicates that 
$b_-$ is located in the left exterior of the horizon. 
 Here, we take the thermofield double state \eqref{TFD} in which states in two exteriors 
are related to each other by Euclidean time evolution over $\beta/2$. 

At $t_b = 0$ or in very early time $t_b \ll r_h$, 
there is no configuration with islands. 
A configuration with an island appears after some time evolution 
but gives only a subdominant saddle in the path integral until the Page time. 
Thus the configuration without the island dominates before the Page time, 
and the entanglement entropy is calculated as 
\begin{equation}
 S = \frac{2\pi b^2}{G_N} 
 + \frac{c}{6} \log \left[\frac{16r_h^2(b-r_h)}{b}\cosh^2\frac{t_b}{2 r_h}\right] \ . 
\end{equation}
In the initial state at $t_b = 0$ gives the minimum of the entanglement entropy 
\begin{equation}
 S = \frac{2\pi b^2}{G_N} 
 + \frac{c}{6} \log \left[\frac{16r_h^2(b-r_h)}{b}\right] \ , 
\end{equation}
and the entanglement entropy continues to increase with time. 
For $t_b\gg r_h$, it is approximated as 
\begin{equation}
 S \simeq \frac{2\pi b^2}{G_N} + \frac{c\,t_b}{6 r_h} \ . 
\end{equation}

After the Page time, the configuration with the island dominates. 
The entanglement entropy is calculated as 
\begin{align}
 S &= \frac{2\pi a^2}{G_{\rm N}} + \frac{2\pi b^2}{G_{\rm N}} 
 + \frac{c}{6} \log \left[\frac{2^8 r_{\rm h}^4(a-r_{\rm h})(b-r_{\rm h})}{ab}\cosh^2\frac{t_a}{2 r_{\rm h}}\cosh^2\frac{t_b}{2 r_{\rm h}}\right] 
\notag\\&\quad
 + \frac{c}{3} \log \left[\frac{\cosh\left(\frac{r_*(a)-r_*(b)}{2r_{\rm h}}\right)-\cosh\left(\frac{t_a-t_b}{2r_{\rm h}}\right)}
 {\cosh\left(\frac{r_*(a)-r_*(b)}{2r_{\rm h}}\right) + \cosh\left(\frac{t_a+t_b}{2r_{\rm h}}\right)}\right] \ , 
\label{totalE}
\end{align}
where 
\begin{equation}
 \cosh\frac{r_*(a)-r_*(b)}{2r_{\rm h}} 
 = 
 \frac{1}{2} \left[\sqrt{\frac{a-r_{\rm h}}{b-r_{\rm h}}}\,e^{ \frac{a-b}{2r_{\rm h}}} + \sqrt{\frac{b-r_{\rm h}}{a-r_{\rm h}}}\,e^{\frac{b-a}{2r_{\rm h}}} \right] \ . 
 \label{abstar}
\end{equation}
The endpoint of the island in the right exterior is indicated by $(t,r) = (t_a,a)$, 
which is determined so that the entanglement entropy is extremized. 
The other endpoint is given by $(t,r) = (- t_a + i \beta/2,a)$, 
since the dominant configuration will be symmetric. 
After the Page time, $t_b$ will be very large, and then, 
the entanglement entropy is extremized for 
\begin{align}
 t_a &= t_b \ , 
 \\
 a 
 &\simeq 
 r_h 
 + \frac{(c\,G_N)^2}{144\pi^2 r_h^2 (b-r_h)} e^{\frac{r_h-b}{r_h}} \ . 
 \label{a-sol}
\end{align}
Then, the entanglement entropy becomes 
\begin{align}
 S 
 &\simeq 
 \frac{2\pi r_h^2}{G_N} + \frac{2\pi b^2}{G_N}
 + \frac{c}{6} 
 \left[
 \log \left(\frac{16 r_h^3 (b-r_h)^2}{b} \right) + \frac{b-r_h}{r_h}  
 \right] \ . 
\end{align}
Thus, the entanglement entropy after the Page time approximately equals to 
twice of the Bekenstein-Hawking entropy. 

Next, we consider the entanglement entropy of the region $R_+$, 
which is interpreted as the entanglement entropy of the Hawking radiation in the right exterior. 
We assume that the island cannot overlap or be related causally to the region $R_-$. 
Then, there is no configuration with islands for the entanglement entropy of $R_+$. 
In order to calculate the entanglement entropy, 
we introduce the IR cut-off of the spacetime, $\Lambda$, in $r\gg r_h$. 
Then, the entanglement entropy is given by 
\begin{equation}
 S
 = 
 \frac{\pi b^2}{G_N} 
 + \frac{c}{6} \log \Lambda \ , 
\end{equation}
namely, the entanglement entropy of $R_+$ is infinitely large. 
This result is consistent with the entanglement entropy for $R$ without the island. 
The entanglement entropy of $R_+$ and $R_-$ must satisfy the subadditivity condition 
with the entanglement entropy of $R = R_+\cup R_-$, 
\begin{equation}
 S(R) \leq S(R_+) + S(R_-) \ . 
\end{equation}
It is always possible to consider the entanglement entropy on the Schwarzschild spacetime 
in models without dynamical gravitation. 
In such models, the islands cannot appear because 
they are nonperturbative saddles of dynamical gravitation. 
The subadditivity condition above should be satisfied only in 
the entanglement entropy without islands. 
The entanglement entropy of the region $R$ without the island 
continues to increase and is not bounded from above. 
In order to satisfy the subadditivity condition, 
the entanglement entropy of the region $R_+$ must be infinitely large. 
This result is the same for models with gravitation 
since there is no configuration with islands for $S(R_+)$. 
It should be also noted that the entanglement entropy of $R_+$ is independent of time 
since the Schwarzschild spacetime and the Hartle-Hawking vacuum is 
invariant under the time evolution as long as we consider 
only one of two exteriors of the horizon. 

Although the entanglement entropy of the Hawking radiation satisfies 
the subadditivity condition, it does not give the minimal output entropy 
which satisfies the additivity condition. 
The entanglement entropy of $R$ has contribution from the island 
and bounded from above by twice of the Bekenstein-Hawking entropy. 
On the other hand, the entanglement entropy of $R_+$ 
has no contribution from the island and is infinitely large. 
Thus, the entanglement entropy of the Hawking radiation always satisfies 
\begin{equation}
 S(R) \ll S(R_+) + S(R_-) \ , 
\end{equation}
and hence it cannot be the minimum output entropy in the additivity conjecture. 

We can consider quantum channels to the black hole $B$, instead 
of the quantum channel to the Hawking radiation. 
The entanglement entropy of $B$ is the complement of the Hawking radiation 
and hence equals to the entanglement entropy of the region $R$. 
The entanglement entropy of the region $B_+$ is different from the entanglement entropy of $R_+$, 
since now the total system is separated into four subsystems $R_+$, $R_-$, $B_+$ and $B_-$. 
The region of $B_+$ is identified with the region between 
the endpoint of $R_+$ and the quantum extremal surface, 
namely, the region between $b_+$ and $a_+$. 
Then, the entanglement entropy of $B_+$ is approximately the same to 
half of the entanglement entropy of $B = B_+ \cup B_-$. 
It is calculated as 
\begin{align}
 S 
 &= 
 \frac{\pi r_h^2}{G_N} + \frac{\pi b^2}{G_N}
 + \frac{c}{12} 
 \left[
 \log \left(\frac{16 r_h^3 (b-r_h)^2}{b} \right) + \frac{b-r_h}{r_h}  
 \right] \ .  
\end{align}
After the Page time, the entanglement entropy of the black hole saturates the subadditivity condition 
and satisfies 
\begin{equation}
 S(B) \simeq S(B_+) + S(B_-) \ , 
\end{equation}
which implies 
\begin{equation}
 S((\mathcal N_+\otimes\mathcal N_-)(\rho)) \simeq S(\mathcal N_+(\rho_+)) + S(\mathcal N_-(\rho_-)) \ . 
\end{equation}
However, the black hole states in the Hartle-Hawking vacuum does not satisfy the additivity conjecture
since the states after the Page time does not give 
the minimum output entropy of $\mathcal N_+\otimes\mathcal N_-$. 
The initial state of $B$ gives smaller entanglement entropy but 
the entanglement entropy of $B_+$ or $B_-$ is independent of time. 
Thus, for the minimum output entropy, we have 
\begin{equation}
 S_\text{min}(\mathcal N_+ \otimes \mathcal N_-) < S_\text{min}(\mathcal N_+) + S_\text{min}(\mathcal N_-) \ , 
\end{equation}
and hence, the Hartle-Hawking vacuum does not give the minimal output entropy in the additivity conjecture. 
 
The argument above does not imply the violation of the additivity conjecture 
since we considered only the Hartle-Hawking vacuum state. 
In order to construct the Hartle-Hawking state from sets of states 
in each exterior of the horizon even approximately, 
we need to prepare non-trivial sets of static states in two exteriors. 
From the sets of states in two sides, we would be able to construct 
another state of the total system. 
Thus, we need to study the entanglement entropy of other states than the Hartle-Hawking vacuum, 
in order to see whether the additivity conjecture has large violation in this setup. 
In the next section, we consider the entanglement entropy of the Boulware vacuum and other static states.


\subsubsection{Entanglement entropy in Boulware and other static vacua}\label{sec:Boulware}

In the Hartle-Hawking vacuum, the geometry can be approximated by the classical Schwarzschild solution. 
In other static vacua, the energy-momentum tensor \eqref{Tuu-vac}--\eqref{Tthth-vac} 
diverges around the Schwarzschild radius, and the geometry is modified from the classical solution 
by quantum corrections as \eqref{qtm-c} and \eqref{qtm-r}. 
The solution of the semi-classical Einstein equation, \eqref{qtm-c} and \eqref{qtm-r}, 
has no event horizon at finite radius for $\kappa\neq \frac{1}{2r_h}$, and hence, 
two exteriors of the horizon are now disconnected each other. 
Thus, the entanglement entropy of the region $R$ 
is simply given by sum of the entanglement entropy of the regions in each exterior $R_+$ and $R_-$, namely, 
\begin{equation}
 S(R) = S(R_+) + S(R_-) \ . 
\end{equation}
Thus, a static vacuum state other than the Hartle-Hawking vacuum 
would be the disentangled states which is proposed in \cite{Hayden:2020vyo}. 
In the following, we calculate the entanglement entropy of the region $R$ 
in the other static vacua than the Hartle-Hawking vacuum. 

The entanglement entropy of the region $R_+$ can 
be calculated by inserting the twist operator at $b_+$. 
Since the geometry has time translation invariance, 
we can put $b_+$ at $(t,r) = (0,b)$ without loss of generality. 
As the geometry is also invariant under the time reversal, 
the island in the most dominant saddle can appear only on the same time slice $t=0$, or equivalently, $U=-V$. 

We consider the static vacuum states which are defined by using the coordinates \eqref{UV}. 
In order to calculate the entanglement entropy in these static vacua, 
we should use the formula \eqref{UV} in the same coordinates. 
The radius $r$ goes to infinity (for $\kappa<\frac{1}{2r_h}$) or zero (for $\kappa>\frac{1}{2r_h}$) 
within a finite distance in coordinates \eqref{UV}, except for $\kappa=0$, 
and hence, it should be treated as the boundary of the geometry.%
\footnote{%
Here, we assume that there is no degree freedom which is localized on the boundary 
since the model has only bulk fields. 
}
Then, we need to introduce the mirror image on the other side of the boundary. 
For $\kappa<\frac{1}{2r_h}$, the radius goes to $r\to\infty$ in $x\to -\infty$, which is $U=V=0$. 
For $\kappa>\frac{1}{2r_h}$, the radius goes to $r=0$ at finite $x$. 
As we will see later, configurations with islands does not dominates for $\kappa>\frac{1}{2r_h}$, 
and then, the position of the boundary can be approximated by $U=V=0$. 

We first calculate the entanglement entropy in the configuration without islands. 
The entanglement entropy is given by the formula \eqref{IslandRule} with \eqref{S2D} 
and depends on the metric only at the positions of the twist operators. 
We put the region $R_+$ sufficiently apart from the horizon, 
and then, the metric can be approximated by the classical solution. 
By using the formulae \eqref{IslandRule} and \eqref{S2D} with 
\eqref{UV}, \eqref{cls-r} and \eqref{W-away}, 
the entanglement entropy of $R_+$ is calculated as 
\begin{align}
 S &= \frac{\pi b^2}{G_N} + \frac{c}{12}\log\left|\frac{UV}{W^2}\right|_{r=b} 
 \notag\\
 &= \frac{\pi b^2}{G_N} + \frac{c}{12}\log\left(\frac{b-r_h}{\kappa^2 b}\right) \ . 
 \label{S-NoIsland-B}
\end{align}
The entanglement entropy depends on the vacuum state and 
increases as $\kappa$ decreases. 
It becomes infinitely large in the Boulware vacuum, $\kappa\to 0$, 
while it is very small for large $\kappa$.%
\footnote{%
The distance between the horizon and $b_+$ in $(U,V)$ coordinates becomes very small 
as $\kappa$ becomes very large, and eventually approaches to the cut-off scale. 
The entanglement entropy becomes zero when the distance is approximately the same to the cut-off scale. 
Then, the energy of the radiation in the vacuum state becomes also comparable to the Planck scale. 
}

Next, we calculate the entanglement entropy in the configuration with an island. 
In the other vacua than the Hartle-Hawking vacuum, two exteriors are disconnected from each other 
and the geometry has another boundary in the black hole side. 
The island extends between the quantum extremal surface and the boundary in the black hole side. 
In the Boulware vacuum, the entanglement entropy 
should be calculated by using the tortoise coordinate $r_*$, 
and the distance to the boundary is infinite in $r_*$. 
However, the entanglement entropy is still finite since 
the region $R_+$ also extends to the spatial infinity, 
and we insert two twist operators at $a_+$ and $b_+$. 
By using \eqref{IslandRule} and \eqref{S2D}, 
the entanglement entropy of $R_+$ is given by  
\begin{equation}
 S 
 = 
 \frac{\pi r^2(x)}{G_N} + \frac{\pi b^2}{G_N} 
 + \frac{c}{12}\log\left|\frac{(U_a(x)-U_b)^2(V_a(x)-V_b)^2(U_a(x)-V_a(x))^2(U_b-V_b)^2}
 {W_a(x)W_b(U_a(x)-V_b)^2(V_a(x)-U_b)^2}\right|
 \label{S-island-B}
\end{equation}
where $U_a$, $V_a$ and $W_a$ are $U$, $V$ and $W$ at $a_+$, 
which are given by \eqref{UV} and \eqref{W-near}, respectively. 
Here, the coordinate $x$ also stands for the position of $a_+$. 
$U_b$, $V_b$ and $W_b$ are $U$, $V$ and $W$ at $b_+$, 
which are given by \eqref{UV}, \eqref{cls-r} and \eqref{W-away}. 
The position of the quamtum extremal surface $a_+$ is given by 
\begin{align}
 0 &= \partial_x S 
 \notag\\
 &\simeq \frac{c}{24 r_h} \left[e^{x/r_h} + 1+ 4\kappa^2 r_h^2\right] 
  - \frac{2c}{3} \kappa \left(\frac{c\,G_N}{48\pi r_h (b-r_h)}\right)^{\kappa r_h} e^{\kappa (x-b+r_h)}
 \notag\\
 &\quad
 + \frac{c^2 G_N}{1152\pi r_h^4} 
    \left(e^{x/r_h} - 1 + 4 \kappa^2 r_h^2\right)
    \left[r_h e^{x/r_h} - (1-4\kappa^2 r_h^2) \left(x + r_h \log\frac{\alpha}{4r_h^2}\right)\right] \ , 
\end{align}
where we have assumed that $\kappa$ is comparable with $1/r_h$. 
The extremum condition above has a solution when $(-x) \gg r_h$ where $e^{x/r_h}$ is negligible. 
The position of the quantum extremal surface is approximately calculated by 
\begin{equation}
 x + r_h \log\left(\frac{\alpha}{4r_h^2}\right) 
 \simeq - \frac{48\pi r_h^3(1 + 4 \kappa^2 r_h^2)}{c\,G_N (1 - 4 \kappa^2 r_h^2)^2} \ . 
 \label{sol-x}
\end{equation}
The quantum extremal surface is located at an inner place 
as $\kappa$ is close to $\frac{1}{2r_h}$, 
and the island does not appear in the Hartle-Hawking vacuum for $t_b = 0$. 
By substituting \eqref{sol-x} to \eqref{S-island-B}, 
the entanglement entropy in the configuration with an island is obtained as 
\begin{align}
 S 
 &= 
 \frac{\pi b^2}{G_N} 
 + \frac{c}{12} \log\left(\frac{b-r_h}{\kappa^4 b}\right)
 + \frac{16\pi r_h^4 \kappa^2 (1 + 12 \kappa^2 r_h^2 -16 \kappa^4 r_h^4 + 64 \kappa^6 r_h^6)}
 {G_N (1 - 4 \kappa^2 r_h^2)^4} \ . 
 \label{S-Island-FK}
\end{align}
The entanglement entropy \eqref{S-Island-FK} is always larger than \eqref{S-NoIsland-B}, and hence, 
the configuration with an island does not becomes the dominate saddle 
as long as $\kappa$ is comparable with $\frac{1}{r_h}$. 

If $\kappa$ is much smaller, the configuration with an island 
dominates over the configuration without islands. 
The result \eqref{S-Island-FK} becomes invalid if $\kappa$ is smaller than $c\,G_N r_h^{-3}$. 
Then, higher order corrections in the small $\kappa$ expansion 
are negligible in the semi-classical approximation, 
and the vacuum state can be approximated by the Boulware vacuum $\kappa=0$. 
The entanglement entropy for $\kappa = 0$ is given by 
\begin{equation}
 S 
 = 
 \frac{\pi r^2(x)}{G_N} + \frac{\pi b^2}{G_N} 
 + \frac{c}{6}\log\left|C^{1/2}(a) C^{1/2}(b)(r_*(a)-r_*(b))^2\right| \ . 
 \label{S-island-Boul}
\end{equation}
The position of the quantum extremal surface is the same as \eqref{sol-x}, 
and then, the entanglement entropy is calculated as 
\begin{equation}
 S = \frac{\pi b^2}{G_N} 
 + \frac{c}{12} \log\left(\frac{b-r_h}{b}\right) + \frac{c}{3}\log\left(\frac{48\pi r_h^3}{c\,G_N}\right) \ , 
\end{equation}
This gives the upper bound of the entanglement entropy for small $\kappa$.


\section{Conclusion and discussions}\label{sec:Conclusion}

In this paper, we have studied the entanglement entropy of the Hawking radiation, 
or equivalently, of the region $R$ in various static vacua in the Schwarzschild spacetime. 
In the Hartle-Hawking vacuum, the entanglement entropy of $R$ is much smaller than 
the sum of the entanglement entropy of $R_+$ and $R_-$. 
When we consider the quantum channel which traces out the black hole states, 
the entanglement entropy in the Hartle-Hawking vacuum does not give 
the minimal output entropy of the additivity conjecture in quantum information theory. 

In the other static vacuum states, in which the energy-momentum tensor 
takes a different asymptotic value from the Hartle-Hawking vacuum, 
the energy-momentum tensor diverges in the classical Schwarzschild solution, 
and hence, solutions of the semi-classical Einstein equation have 
quite different structure around the schwarzschild radius 
due to the back-reaction from the vacuum energy. 
The geometry has no horizon, and two exteriors are disconnected with each other. 
Then, the entanglement entropy $S(R)$ is simply given by sum of $S(R_+)$ and $S(R_-)$. 
The entanglement entropy becomes larger for smaller vacuum energy 
and becomes smaller as the radiation in the vacuum state becomes stronger. 
The entanglement entropy is bounded from above due to effects of the island. 

Although we have not proven that the additivity conjecture is satisfied 
by quantum states around the Schwarzschild black hole, 
the results in this paper implies that the additivity conjecture is satisfied. 
In order to construct the Hartle-Hawking vacuum, 
we need to prepare sets of states in both exteriors of the horizon. 
Then, some of the other static vacuum states 
can also be constructed from the sets of states, 
even if the sets of states are minimal to form the Hartle-Hawking states approximately. 
Although these states are sometimes excluded as unphysical states, 
they should be taken into account for the additivity conjecture. 
Then, by taking quantum effects into consideration, 
two exteriors are disconnected with each other. 
The quantum effects modifies the geometry near the Schwarzschild radius 
even if the deviation from the Hartle-Hawking vacuum is very small, 
in which the entanglement entropy is approximately the same to 
the minimal entropy in the Hartle-Hawking vacuum at $t=0$. 
Thus, the violation of the additivity conjecture would be very small even if it exists. 

Although we studied only static vacuum states for simplicity, 
it is straightforward to generalize to time dependent states. 
For example, the most suitable vacuum state for black holes 
which are formed by gravitational collapses would be the Unruh vacuum. 
The Unruh vacuum has the same outgoing Hawking radiation to the Hartle-Hawking vacuum, 
but has no incoming radiation as in the Boulware vacuum. 
If we consider the Unruh vacuum in one exterior of the classical Schwarzschild geometry, 
the energy-momentum tensor has divergence at the past horizon, namely $V=0$ in the Kruskal coordinate,  
and the divergence continues to the future horizon of the other exterior. 
The semi-classical solution is disconnected at $V=0$ due to the quantum effects, 
and $r\to\infty$ at $V=0$ in this case. 
The singularity appears at $V=0$ since the solution is a vacuum solution, 
which has no classical matter in the bulk. 
The matters are located at an exceptional point, which is nothing but the singularity at $V=0$. 
In a more realistic configuration, matters are distributed in a finite region, 
and then, the singularity is smoothed out. 
The semi-classical geometry in the Unruh vacuum can be obtained 
by introducing slow time evolution by the Hawking radiation 
into the semi-classical geometry in the Boulware vacuum \cite{Ho:2018jkm,Ho:2019pjr}. 

It should be noted that the Schwarzschild geometries in this paper 
is obviously different from the geometry of real black holes. 
Black holes in our universe would be formed by the gravitational collapse of matters, 
and hence, neither the Einstein-Rosen bridge nor naked singularity would exist in real black holes. 
In real black hole geometries, the interior should be replaced by another solution with matters. 
The Einstein-Rosen bridge and singularity would be theoretical replacements 
to reproduce the entanglement and the energy of the collapsed matters. 

In this paper, we have studied only the four-dimensional Schwarzschild spacetime, 
but generalization to other black holes is straightforward. 
The RST model is solvable, and black hole solutions in the other static vacua can be obtained, exactly. 
In particular, the conical singularity in $\kappa>\frac{1}{2r_h}$ would be absent in the RST model 
since two-dimensional models have no angular directions. 
In the framework of the AdS/CFT correspondence, it would be more interesting 
to study the entanglement wedge of the black hole, which corresponds to CFT. 
The island implies that only the geometry outside the quantum extremal surface 
can be reconstructed from CFT. 
Thus, if we consider the Boulware vacuum in the AdS black hole geometries, 
the region $r\gg r_h$ near the black hole cannot be reproduced by CFT. 
Generalizations to other models are left for future studies.

\section*{Acknowledgments}

The author would like to thank Koji~Hashimoto for valuable comments. 
This work is supported in part by JSPS KAKENHI Grant No.~JP17H06462 and JP20K03930.

\begin{appendix}

\section{Semiclassical Schwarzschild geometry in static vacua}\label{sec:nh}

In this section, we derive the semi-classical Schwarzschild geometry near the Schwarzschild radius. 
We solve the semi-classical Einstein equation \eqref{SemiEin}, 
where $\langle T_{\mu\nu}\rangle$ is given by \eqref{Tuu-vac}--\eqref{Tthth-vac}. 
The most general spherically symmetric metric is given by \eqref{metric0}, 
up to the coordinate transformation.  
In terms of the $(u,v)$ coordinates \eqref{uv}, 
the metric components depend only on the tortoise coordinates $r_*$, namely, 
\begin{equation}
 ds^2 = - C(r_*) \left(dt^2 - dr_*^2\right) + r(r_*)^2 d\Omega^2 \ . 
\end{equation}
We solve the semi-classical Einstein equation \eqref{SemiEin} in static vacua \eqref{F-vac}. 

The energy-momentum tensor on the classical Schwarzschild geometry diverges at the event horizon. 
This implies that quantum corrections are negligible away from the black hole 
but become important near the Schwarzschild radius. 
In terms of the tortoise coordinates, the Schwarzscihld metric \eqref{MetricS} 
is given by \eqref{cls-c} and \eqref{cls-r}, which behaves near the Schwarzschild radius as 
\begin{align}
 C(r_*) &= \mathcal O(G_N) \ , 
 & 
 r(r_*) &= r_h + \mathcal O(G_N) \ . 
 \label{NH}
\end{align}
Since quantum effects do not modify the classical geometry away from the Schwarzschild radius, 
we expect that the metric still behaves as \eqref{NH} even taking quantum effects into account, 
but $\mathcal O(G_N)$ terms in \eqref{NH} would be different from the classical solution 
\eqref{cls-c-nh} and \eqref{cls-r-nh}. 
Thus, we calculate $\mathcal O(G_N)$ terms by solving the semi-classical Einstein equation \eqref{SemiEin}. 
we consider the following expansion; 
\begin{align}
 C(x) &= G_N C_1(x) + \mathcal O(G_N^2) \ , 
 \\
 r(x) &= r_h + G_N r_1(x) + \mathcal O(G_N^2) \ , 
\end{align}
where $x$ is the tortoise coordinate near the horizon \eqref{x}. 
At the leading order, the semi-classical Einstein equation is 
\begin{align}
 0 &= 
 - 4 \alpha \kappa^2 C_1^2(x) - 4 C_1^3(x) + \alpha C_1^{\prime\,2}(x) + 4 r_h r_1'(x) C_1(x) C_1'(x) \ , 
 \label{eq1}
 \\
 0 &= 
 - 2 C_1^3(x) - \alpha C_1^{\prime\,2}(x) + \alpha C_1(x) C_1''(x) + 2 r_h r_1''(x) C_1^2(x) \ , 
 \label{eq2}
\end{align}
where $\alpha$ is defined by \eqref{alpha}. 

From \eqref{eq1}, we obtain 
\begin{equation}
 r_1(x) 
 = \int dx 
 \frac{4 \kappa^2 C_1^2(x) + 4 C_1^3(x) - \alpha C_1^{\prime\,2}(x)}{2 r_h C_1(x) C_1'(x)} \ . 
\end{equation}
By substituting this solution, \eqref{eq2} becomes 
\begin{equation}
 0 = 
 \left(4 \kappa^2 C_1^2(x) + 4 C_1^3(x) - \alpha C_1^{\prime\,2}(x)\right)
 \left(- C_1^{\prime\,2}(x) + C_1(x) C_1''(x)\right) \ . 
\end{equation}
The solution is given by 
\begin{align}
 C_1(x) &= e^{\lambda (x - x_0)} \ , \qquad \frac{\alpha \kappa^2}{\sinh^2[\kappa(x-x_0)]} \ , 
\end{align}
where $\lambda$ and $x_0$ are integration constants. 
Thus, we have obtained  
\begin{align}
 C_1(x) &= e^{\lambda (x - x_0)} \ , 
 & 
 r_1(x) &= 
 \frac{1}{\lambda^2 r_h} e^{\lambda (x - x_0)}  
 - \frac{\alpha}{4 r_h}\left(\lambda - \frac{4 \kappa^2}{\lambda}\right) \left(x - x_0\right) \ , 
 \label{sol}
\end{align}
or 
\begin{align}
 C_1(x) &= \frac{\alpha \kappa^2}{\sinh^2[\kappa(x-x_0)]} \ , 
 & 
 r_1(x) &= 0 \ . 
 \label{sol-f}
\end{align}
Here, the solution \eqref{sol-f} is an exact solution of the semi-classical Einstein equation 
\eqref{SemiEin} with \eqref{Tuu-vac}--\eqref{Tthth-vac} 
but is unphysical because the radius is exactly constant. 
Thus, the solution \eqref{sol} is the leading order solution of 
the semi-classical Einstein equation near the Schwarzschild radius. 
The integration constants are determined by 
the junction conditions with the Schwarzschild solution outside the near horizon region as 
\begin{align}
 \lambda &= \frac{1}{r_h} \ , 
 & 
 x_0 &= - r_h \log\left(\frac{\alpha}{4r_h^2}\right) \ . 
\end{align}
Thus, we obtain the semi-classical Schwarzschild geometry near the horizon, 
\eqref{qtm-c} and \eqref{qtm-r}. 

\end{appendix}


\end{document}